\documentclass[%
reprint,
superscriptaddress,
showpacs,preprintnumbers,
 amsmath,amssymb,
 aps,
prd,
]{revtex4-1}

\usepackage{graphicx}
\usepackage{dcolumn}
\usepackage{bm}
\usepackage{bbold}
\usepackage{amssymb,amsmath}
\usepackage{hyperref}

\usepackage{color}
\usepackage{amsfonts}
\usepackage{subfigure}
\usepackage{array}

\newcommand{\Tr}{\ensuremath{\operatorname{Tr}}}
\newcommand{\tr}{\ensuremath{\operatorname{tr}}}

\newcolumntype{L}{>{\centering\arraybackslash}m{3cm}}

\definecolor{bjcol}{rgb}{1,.44,0.13}

\definecolor{blue}{rgb}{0,0,1}

\definecolor{green}{rgb}{0,1,0}

\definecolor{red}{rgb}{1,0,0}

\definecolor{gray}{rgb}{.5,.5,.5}

\definecolor{darkgreen}{rgb}{.0,.5,.0}

\def\Fig#1{Fig.~\ref{#1}}

\def\Eq#1{Eq.~(\ref{#1})}

\def\eqref#1{(\ref{#1})}

\def\tab#1{Tab.~\ref{#1}}

\def\sec#1{Sec.~\ref{#1}}

\def\lA0{{\langle A_0 \rangle}}
\def\bA0{{\bar{A}_0}}

\def\0#1#2{\frac{#1}{#2}}

\graphicspath{{./figures/}{./}}

\begin{document}

\preprint{}

\title{Correlations of conserved charges and QCD phase structure}

\author{Rui Wen}
\affiliation{School of Physics, Dalian University of Technology, Dalian, 116024,
  P.R. China}

\author{Wei-jie Fu}
\email{wjfu@dlut.edu.cn}
\affiliation{School of Physics, Dalian University of Technology, Dalian, 116024,
  P.R. China}


\begin{abstract}

Correlations of conserved charges, i.e., the baryon number, electric charge and the strangeness, have been calculated at finite temperature and chemical potential up to the fourth order. The calculations are done in a 2+1 flavor low energy effective theory, where quantum and thermal fluctuations are encoded through the evolution of flow equations within the functional renormalization group approach. Strangeness neutrality and a fixed ratio of the electric charge to the baryon number density are implemented throughout the computation. We find that higher-order correlations carry more sensitive critical dynamics in comparison to the quadratic ones, and a non-monotonic dependence of the fourth-order correlations between the baryon number and strangeness, $-\chi^{BS}_{31}/\chi^{S}_{2}$ and $\chi^{BS}_{22}/\chi^{S}_{2}$, on the collision energy is found.

\end{abstract}

\pacs{11.30.Rd, 
          11.10.Wx, 
          05.10.Cc, 
          12.38.Mh  
     }                             
\maketitle


\section{Introduction}
\label{sec:intro}

Studies of the QCD phase structure have attracted a number of attentions in the past decade, and enormous efforts, both from the experimental and theoretical sides, were involved in the promising but challenging task to search for the critical end point (CEP) in the QCD phase diagram. The first principle lattice QCD simulations indicate that the finite-temperature QCD phase transition is a continuous analytic crossover in the low baryon chemical potential or density regime \cite{Aoki:2006we,Bellwied:2015rza,Bazavov:2017dus,Bazavov:2017tot,Bazavov:2018mes,Ding:2019prx}. Although the predictive capacity of the lattice simulations is hampered by the sign disaster when the baryon chemical potential is high, other complementary first principle theoretical methods, e.g. the widely used functional approaches of QCD, like the functional renormalization group (fRG) \cite{Braun:2008pi,Braun:2009gm,Mitter:2014wpa,Braun:2014ata,Cyrol:2016tym,Cyrol:2017ewj,Cyrol:2017qkl,Fu:2019hdw} and Dyson-Schwinger equations (DSE) \cite{Fischer:2011mz,Fischer:2014ata,Gao:2015kea,Gao:2016qkh,Fischer:2018sdj,Isserstedt:2019pgx}, have predicted the existence of a CEP in the phase diagram spanned by the temperature and the baryon chemical potential. Note, however, that the existence of the CEP, and if it exists, its location in the phase diagram, are still open questions.

In the past decade, the Beam Energy Scan (BES) Program at the Relativistic Heavy Ion Collider (RHIC) has made significant progress in the experimental studies of the QCD phase structure. Moments of the net-proton multiplicity distributions of different orders were measured, and a non-monotonic dependence of the kurtosis of the net-proton distributions on the beam energy was found \cite{Adamczyk:2013dal,Luo:2015ewa}. Moreover, moments of the net-charge multiplicity distributions \cite{Adamczyk:2014fia} and those of net-kaon multiplicity distributions \cite{Adamczyk:2017wsl} were also measured, for more details, see e.g. \cite{Luo:2017faz,Adamczyk:2017iwn} and references therein. Recently, the measurements of the second-order off-diagonal cumulants, i.e., the correlations of net-proton, net-charge and net-kaon multiplicity distributions have been reported by the STAR Collaboration \cite{Adam:2019xmk}. In response to the experimental measurements, in this work we would like to investigate the correlations of conserved charges, viz. the baryon number, electric charge and the strangeness from the theoretical side. The order of the correlations is not just constrained to be the quadratic, but also extended to the more interesting higher-order ones up to the fourth. The dependence of various correlations on the temperature and baryon chemical potential in the strangeness neutral system with a fixed value of the ratio between the electric charge to baryon number density will be addressed. Furthermore, we will also study the evolution of these correlations with the beam energy in the heavy ion collision experiments. After an extensive studies of all the correlations of conserved charges up to the fourth order, we would like to pick several ones that demonstrate the most non-monotonic behavior, and thus are potentially useful in the experiments.

We employ the 2+1 flavor low energy effective theory within the fRG approach, which has already been used and described in detail in our former work \cite{Wen:2018nkn}. In the fRG approach, quantum fluctuations of different momentum scales are encoded successively through the evolution of the renormalization group (RG) scale from the ultraviolet (UV) to infrared (IR) regimes \cite{Wetterich:1992yh}, so the fRG approach is very suitable for the description of physical systems, which is featured with a distinct hierarchy of scales, and thus entail transformation of the effective degrees of freedom. QCD just belongs to this kind of systems. Recent years have seen significant progress in the first principle QCD studies with the fRG approach \cite{Mitter:2014wpa,Braun:2014ata,Cyrol:2016tym,Cyrol:2017ewj,Cyrol:2017qkl,Fu:2019hdw}. It is remarkable to note that, very recently a phase diagram has been extracted from a detailed QCD calculation at nonzero temperature and density within the fRG approach, and a critical end point is predicted in the phase diagram \cite{Fu:2019hdw}. Moreover, the fRG is also widely employed in low energy effective models, see e.g. \cite{Schaefer:2004en,Schaefer:2006ds,Herbst:2010rf,Skokov:2010wb,Braun:2011iz,Fukushima:2012xw,Haas:2013qwp,Herbst:2013ufa,Tripolt:2013zfa,Fu:2015naa,Fu:2015amv,Fu:2016tey,Rennecke:2016tkm,Jung:2016yxl,Braun:2017srn,Braun:2018svj,Fu:2018qsk,Fu:2018swz,Sun:2018ozp,Wen:2018nkn,Yin:2019ebz} for a selective list of relevant work, and see e.g. \cite{Berges:2000ew,Pawlowski:2005xe,Schaefer:2006sr,Gies:2006wv,Braun:2011pp,Pawlowski:2014aha} for reviews on the fRG approach.

This paper is organized as follows. In \sec{sec:PQM} we briefly introduce the 2+1 flavor low energy effective theory within the fRG approach. In \sec{sec:Thermo} we discuss the thermodynamics and the generalized susceptibilities at finite temperature and chemical potentials. The two constraints, i.e., the strangeness neutrality and the fixed ratio of the electric charge to baryon number density, are implemented. The resulting equilibrium strangeness and electric charge chemical potentials are investigated, and are also expanded in powers of the baryon chemical potential, which can be compared with the relevant lattice results.  In \sec{sec:correlation} the correlations of conserved charges, viz. the baryon number, electric charge and the strangeness, are calculated at finite temperature and density up to the fourth order. The dependence of various correlations on the collision energy will also be discussed. In \sec{sec:sum} we give a summary with discussions.

\section{2+1 flavor low energy effective theory}
\label{sec:PQM}

In this work we employ the Polyakov-loop improved quark-meson theory with $N_f=2+1$ flavor quarks, which was also used in our former work \cite{Wen:2018nkn}. Here we give a brief introduction, and more details about the theory and our calculations can be found in the reference above. The scale dependent effective action is given by
\begin{align}
	\Gamma_{k}[\Phi]=&\int_x \bigg\{ \bar{q} [\gamma_\mu \partial_\mu-\gamma_0(\hat\mu+ig A_0)]q+h\,\bar{q} \,\Sigma_5 q\nonumber\\[2ex]
&+\tr(\bar D_\mu \Sigma \cdot \bar D_\mu\Sigma^\dagger)+\tilde{U}_{k}(\Sigma)+V_{\text{\tiny{glue}}}(L, \bar L)\bigg\}\,,\label{eq:action}
\end{align}
with $\int_{x}\equiv\int_0^{1/T}d x_0 \int d^3 x$ and $\Phi=(q, \bar{q}, \sigma^a, \pi^a)$ ($a=0,1,\cdots,8$), where $q$, $\bar{q}$, $\sigma^a$, and $\pi^a$ are the quark, antiquark, scalar and pseudo-scalar mesonic nonets, respectively. $\hat{\mu}=\mathrm{diag}(\mu_u,\mu_d,\mu_s)$ is the matrix of the quark chemical potentials in the flavor space. The kinetics of mesons is encoded within the trace term in \Eq{eq:action}, where the covariant derivative reads
\begin{align}
  \bar D_\mu \Sigma=\partial_\mu \Sigma +\big[\delta_{\mu 0} \hat \mu ,\Sigma\big]\,,
\end{align}
with $\Sigma=T^a(\sigma^a+i \pi^a)$, viz. the mesonic fields are in the adjoint representation of $U_{V}(3)\times U_{A}(3)$. Here $T^a$ are the generators of the $U(N_f)$ group with $\tr(T^aT^b)=\delta^{ab}/2$. Quarks and mesons couple with each other through the Yukawa coupling with $\Sigma_5=T^a(\sigma^a+i \gamma_5\pi^a)$. The meson masses and the interactions among them are governed by the mesonic potential, which reads
\begin{align}
  \tilde{U}_{k}(\Sigma)&=U_k(\rho_1,\tilde{\rho}_2)-c_A \xi-j_L\sigma_L-j_S\sigma_S\,, \label{eq:tildeU}
\end{align}
with 
\begin{align}
  \rho_1&=\text{tr}(\Sigma \cdot \Sigma^\dagger)\,, \label{eq:rho1}\\[2ex]
  \tilde{\rho}_2&=\text{tr}\Big(\Sigma \cdot \Sigma^\dagger-\frac{1}{3}\,\rho_1\,\mathbb{1}_{3\times 3}\Big)^2 \,,\label{eq:rho2}
\end{align}
and 
\begin{align}
  \xi&=\det(\Sigma)+\det(\Sigma^\dagger)\,, \label{}
\end{align}
where $\rho_1$ and $\tilde{\rho}_2$ are invariant under the transformation of $U_{V}(3)\times U_{A}(3)$, and the Kobayashi-Maskawa-'t Hooft determinant $\xi$ breaks the $U_{A}(1)$ symmetry and keeps the remaining ones unchanged. The two $j_{L/S}$ terms in \Eq{eq:tildeU} break the chiral symmetry explicitly, and are related to the light, strange current quark mass, respectively. Here we have used the light-strange quark basis implicitly, that is related to the singlet-octet basis through a proper rotation as follows,
\begin{align}
  \begin{pmatrix} \sigma_L \\ \sigma_S \end{pmatrix}
  &=\frac{1}{\sqrt{3}}\begin{pmatrix} 1 & \sqrt{2}\\ -\sqrt{2} & 1 \end{pmatrix}
  \begin{pmatrix}\sigma_8\\ \sigma_0 \end{pmatrix}\,. \label{}
\end{align}

The quark confinement and its relevant deconfinement phase transition are encoded, in the statistical sense, through the temporal component of the gluon background field $\langle A_0 \rangle$ on the r.h.s. of \Eq{eq:action}, or the related Polyakov loops, i.e.,
$L(\bm{x})=\langle \Tr\, {\cal P}(\bm x)\rangle/N_c$ and $\bar L (\bm{x})=\langle
  \Tr\,{\cal P}^{\dagger}(\bm x)\rangle/N_c$ with $\langle\cdots \rangle$ denoting the ensemble average. Here one has 
\begin{align}
  {\cal P}(\bm x)= \mathcal{P}\exp\Big(ig\int_0^{\beta}d\tau A_0(\bm{x},\tau)\Big)\,,\label{eq:Ploop}
\end{align}
with the path ordering operator $\mathcal{P}$ on the r.h.s. For more details about the Polyakov loop and its application in the phenomenology, see, e.g. \cite{Fukushima:2003fw,Ratti:2005jh,Schaefer:2007pw,Fu:2007xc}. The dynamics of Polyakov loops is governed by the glue potential $V_{\text{\tiny{glue}}}(L, \bar L)$ in the effective action. In this paper we employ the parameterization of $V_{\text{\tiny{glue}}}$ with the $SU(N_c)$ Haar measure \cite{Lo:2013hla}, which reads
\begin{align}
  \bar V_{\text{\tiny{glue}}}(L,\bar L)=& -\frac{a(T)}{2} \bar L L + b(T)\ln M_H(L,\bar{L})\nonumber \\[2ex]
  &\quad + \frac{c(T)}{2} (L^3+\bar L^3) + d(T) (\bar{L} L)^2\,,\label{eq:GluepHaar}
\end{align}
with the dimensionless $\bar V_{\text{\tiny{glue}}}=V_{\text{\tiny{glue}}}/T^4$ and the Haar measure as follows
\begin{align}
M_H (L, \bar{L})&= 1 -6 \bar{L}L + 4 (L^3+\bar{L}^3) - 3  (\bar{L}L)^2\,.
\end{align}
The temperature dependence of coefficients in \Eq{eq:GluepHaar} are given by
\begin{align}
  x(T) &= \frac{x_1 + x_2/(t_r+1) + x_3/(t_r+1)^2}{1 + x_4/(t_r+1) + x_5/(t_r+1)^2}\,,\label{eq:xT}
\end{align}
for $x\in \{a, c, d\}$, and 
\begin{align}
  b(T) &=b_1 (t_r+1)^{-b_4}\left (1 -e^{b_2/(t_r+1)^{b_3}} \right)\,,\label{eq:bT}
\end{align}
where $t_r=(T-T_c)/T_c$ is the reduced temperature. The parameters in \Eq{eq:xT} and \Eq{eq:bT} are fixed by fitting the 
thermal behaviors of the Polyakov loop, including its fluctuations, and the thermodynamics in the Yang-Mills (YM) theory at finite temperature, and see \cite{Lo:2013hla} for their values. The back reaction of the quarks on the glue potential can be well captured through an appropriate rescale for the reduced temperature \cite{Haas:2013qwp}, to wit, 
\begin{align}
  (t_r)_{\text{\tiny{YM}}}&\rightarrow \alpha\,(t_r)_{\text{\tiny{glue}}},\quad \text{with} \quad (t_r)_{\text{\tiny{glue}}}=(T-T_c^\text{\tiny{glue}})/T_c^\text{\tiny{glue}}\,.\label{}
\end{align}
In our former work, we have investigated the influence of the two parameters $\alpha$ and $T_c^\text{\tiny{glue}}$ on the QCD thermodynamics in details, see Fig. 9 and Fig. 10 in \cite{Wen:2018nkn}. In this work we adopt $\alpha=0.52$ and $T_c^\text{\tiny{glue}}=270$ MeV, which shows the best agreement with the lattice calculations.

The evolution of the scale dependent effective action in \Eq{eq:action} is described by the Wetterich equation \cite{Wetterich:1992yh}, which reads
\begin{align}
  \partial_{t}\Gamma_{k}&=-\mathrm{Tr}\big(G^{q\bar q}_{k}\partial_{t} R^{q}_{k}\big)+\frac{1}{2}\mathrm{Tr}\big(G^{\phi\phi}_{k}\partial_{t} R^{\phi}_{k}\big)\,,
 \label{eq:WetterichEqPQM}
\end{align}
with $t=\ln (k/\Lambda)$, and here $\Lambda$ is the ultraviolet cutoff of the effective theory, i.e., the initial evolution scale for the flow equation. $G^{q\bar q}_{k}$ and $G^{\phi\phi}_{k}$ are the propagators for the quarks and mesons, respectively. In this work we use the $3d$ flat infrared regulators $R^{q}_{k}$ and $R^{\phi}_{k}$, which are suitable for computations at finite temperature and densities, and the explicit expressions of the regulators can be found in, e.g., Appendix A in \cite{Wen:2018nkn}. Inserting the effective action in \Eq{eq:action} into the flow equation in \Eq{eq:WetterichEqPQM}, one is led to the flow equation for the effective potential $\tilde{U}_{k}$ in \Eq{eq:action}, or $U_k$ in \Eq{eq:tildeU}. Obviously, the effective potential is the only term which is scale dependent in \Eq{eq:action}, and this kind of truncation is also called as the local potential approximation (LPA). Within LPA one is left with only the flow equation of $\tilde{U}_{k}$. After the flow is evolved from the UV cutoff toward the IR limit, i.e., $k=0$, quantum fluctuations of different scales, as well as thermal and density fluctuations, are encoded in the effective potential $\tilde{U}_{k=0}$. For more details about solving the flow equation and relevant numerical settings, we refer to our former work \cite{Wen:2018nkn}. Finally, we obtain the thermodynamical potential density, which reads
\begin{align}
  \Omega=\left(\tilde{U}_{k=0}(\sigma_L,\sigma_S)+V_{\text{\tiny{glue}}}(L, \bar L)\right)\bigg|_{\mathrm{\tiny{EoM}}}\,,\label{}
\end{align}
where the subscript EoM denotes that the $\sigma_L$, $\sigma_S$, $L$ and $\bar L$ are on their respective equations of motion. Thus the pressure is given by
\begin{align}
  p=-\Omega\,,\label{eq:pressure}
\end{align}

\section{Thermodynamics at finite chemical potentials}
\label{sec:Thermo}

%
\begin{figure*}[t]
\includegraphics[width=1.\textwidth]{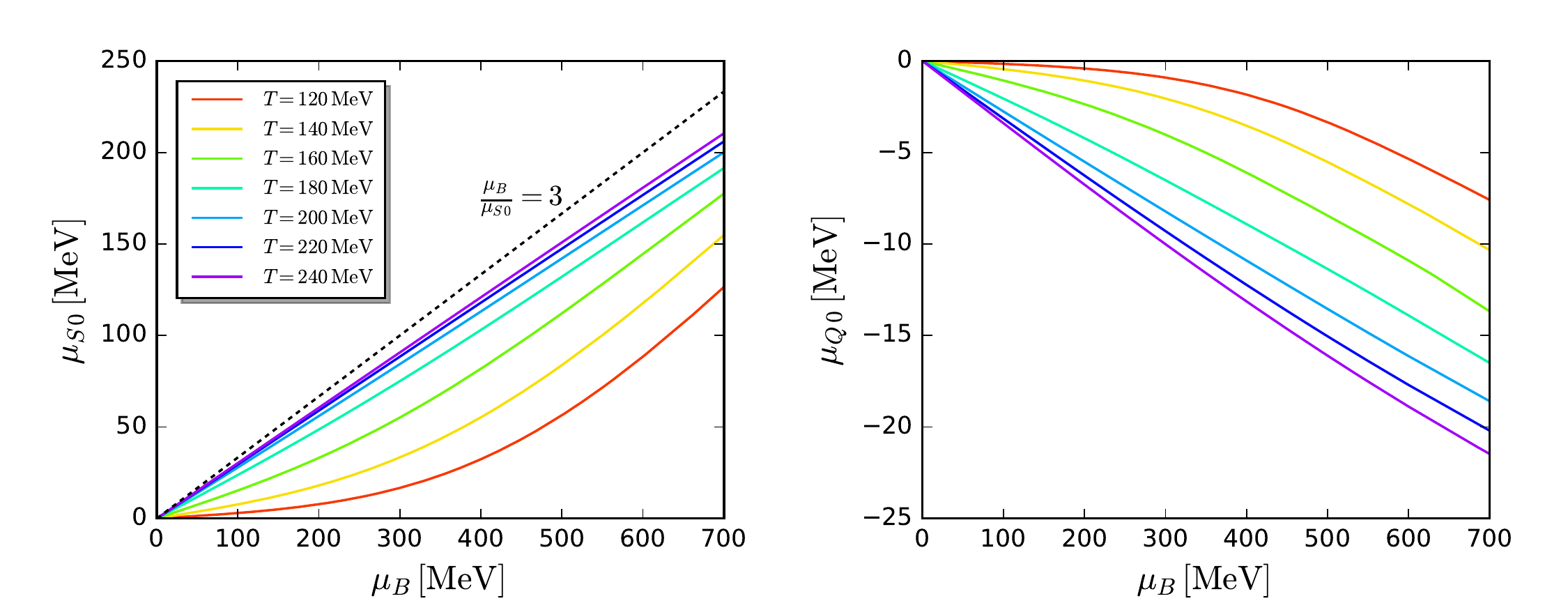}
\caption{Strangeness chemical potential ${\mu_{S}}_0$ (left panel) and electric charge chemical potential ${\mu_Q}_0$ (right panel) as functions of the baryon chemical potential $\mu_B$ with $r=n_Q/n_B=0.4$ and $n_S=0$, for several different values of the temperature. In the left panel we also show the line ${\mu_{S}}_0=\mu_B/3$ (back dashed).}\label{fig:muSQ0}
\end{figure*}
%

%
\begin{figure*}[t]
\includegraphics[width=1.\textwidth]{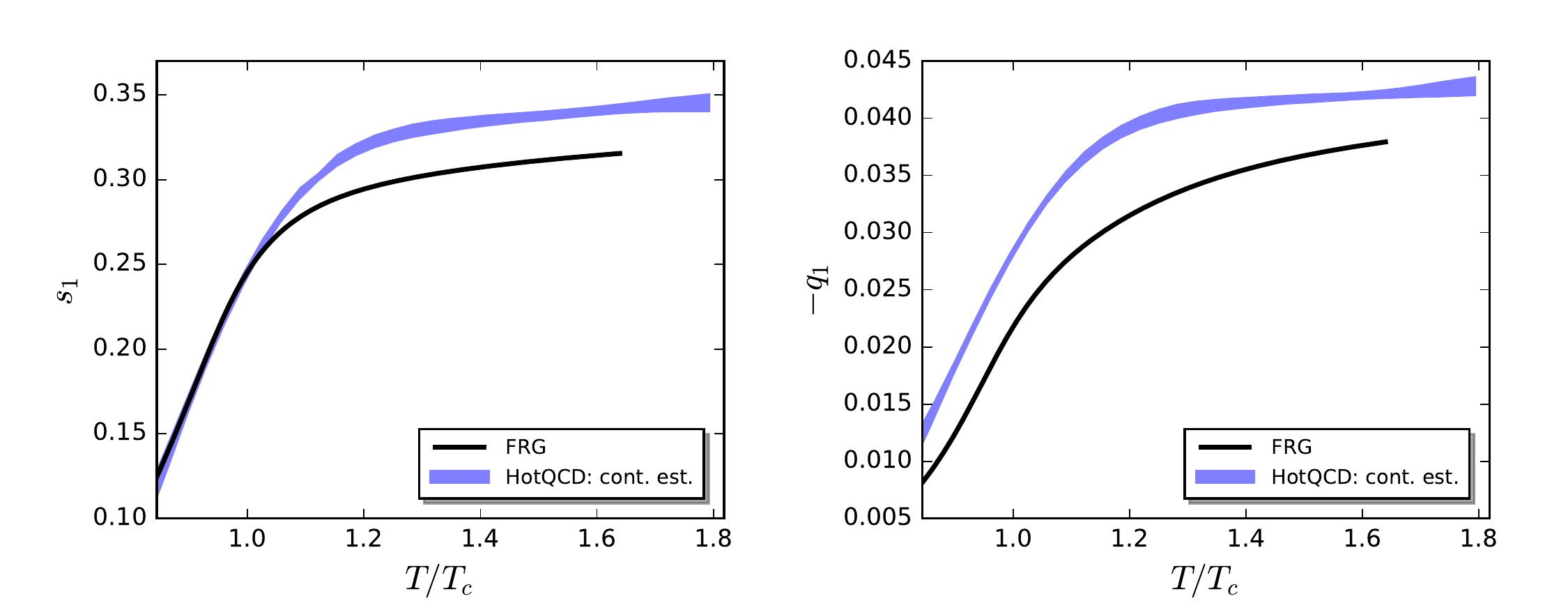}
\caption{Leading order expansion coefficients of the strangeness and electric charge chemical potentials in powers of $\hat \mu_B$, i.e.,  $s_1$ (left panel) and $-q_1$ (right panel) in Eqs.(\ref{eq:muS0}) and (\ref{eq:muQ0}), as functions of the temperature in units of $T_c$, in comparison to the relevant lattice results in \cite{Bazavov:2017dus} as depicted by the blue bands. $T_c$ is the pseudo-critical temperature of the chiral crossover at vanishing chemical potentials. The two constraints $r=n_Q/n_B=0.4$ and $n_S=0$ have been implemented in the calculations.}\label{fig:S1Q1}
\end{figure*}
%

%
\begin{figure*}[t]
\includegraphics[width=1.\textwidth]{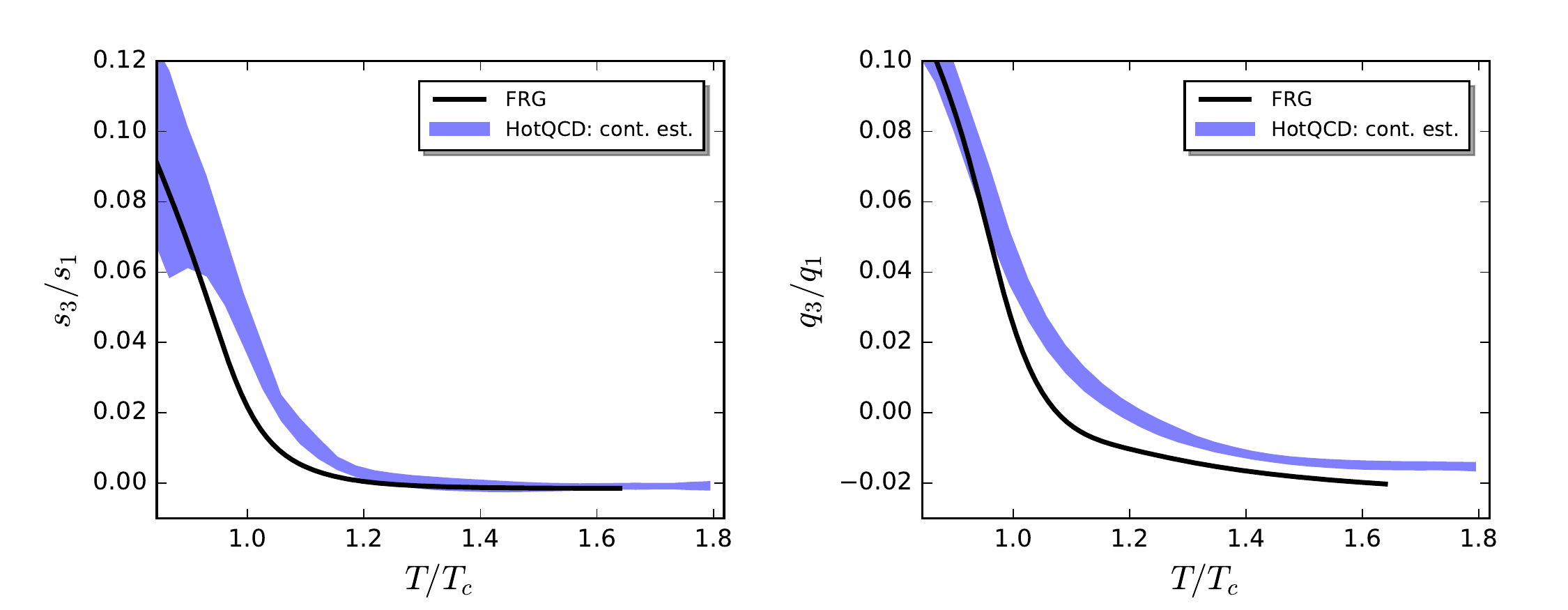}
\caption{$s_3/s_1$ (left panel) and $q_3/q_1$ (right panel), see Eqs.(\ref{eq:muS0}) and (\ref{eq:muQ0}), as functions of the temperature in units of $T_c$, in comparison to the relevant lattice results in \cite{Bazavov:2017dus} as depicted by the blue bands. $T_c$ is the pseudo-critical temperature of the chiral crossover at vanishing chemical potentials. The two constraints $r=n_Q/n_B=0.4$ and $n_S=0$ have been implemented in the calculations.}\label{fig:S31Q31}
\end{figure*}
%

%
\begin{figure*}[t]
\includegraphics[width=1.\textwidth]{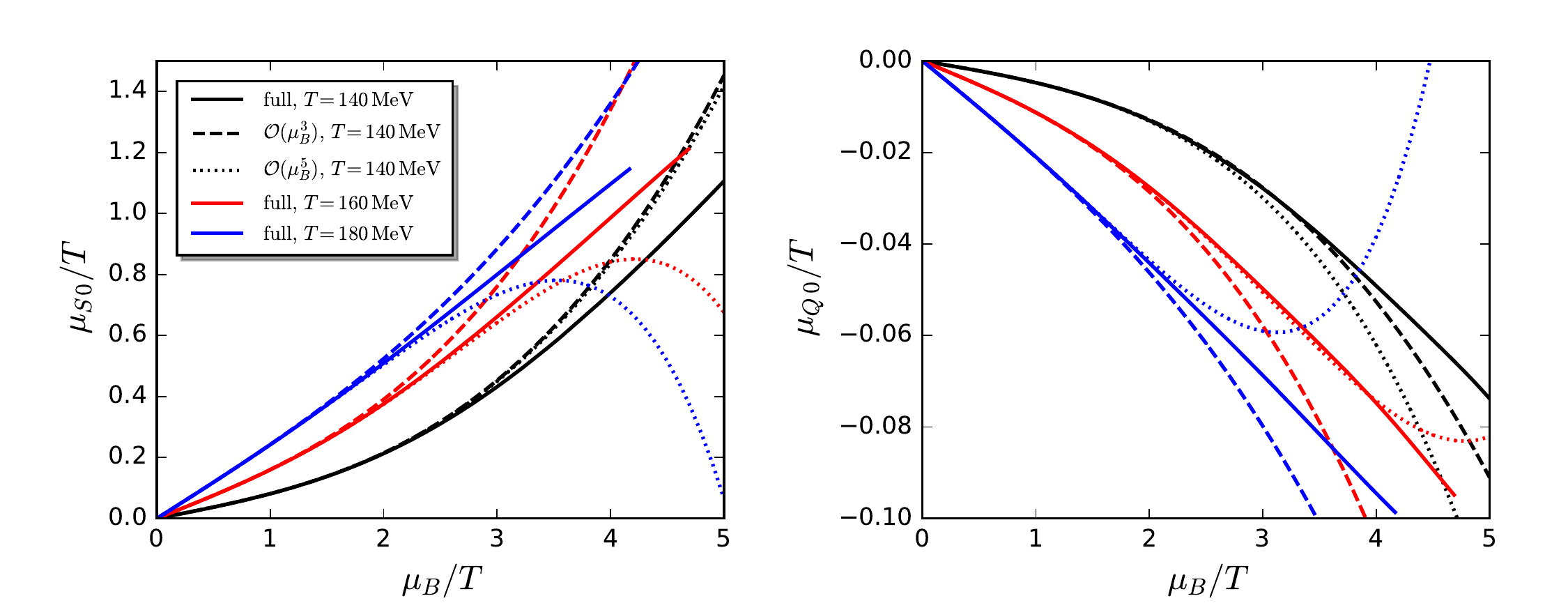}
\caption{Strangeness chemical potential ${\mu_{S}}_0$ (left panel) and electric charge chemical potential ${\mu_Q}_0$ (right panel) as functions of the baryon chemical potential $\mu_B$ normalized by $T$ with $r=n_Q/n_B=0.4$ and $n_S=0$, obtained from Eqs. (\ref{eq:dmuSdmuB}) and (\ref{eq:dmuQdmuB}) (solid lines) in comparison to those from the Taylor expansion in Eqs. (\ref{eq:muS0}) and (\ref{eq:muQ0}) up to the orders of $\mathcal{O}(\mu_B^3)$ (dashed lines) and $\mathcal{O}(\mu_B^5)$ (dotted lines). Different colors are employed to distinguish different values of the temperature.}\label{fig:muSQcompa}
\end{figure*}
%

%
\begin{figure*}[t]
\includegraphics[width=1.\textwidth]{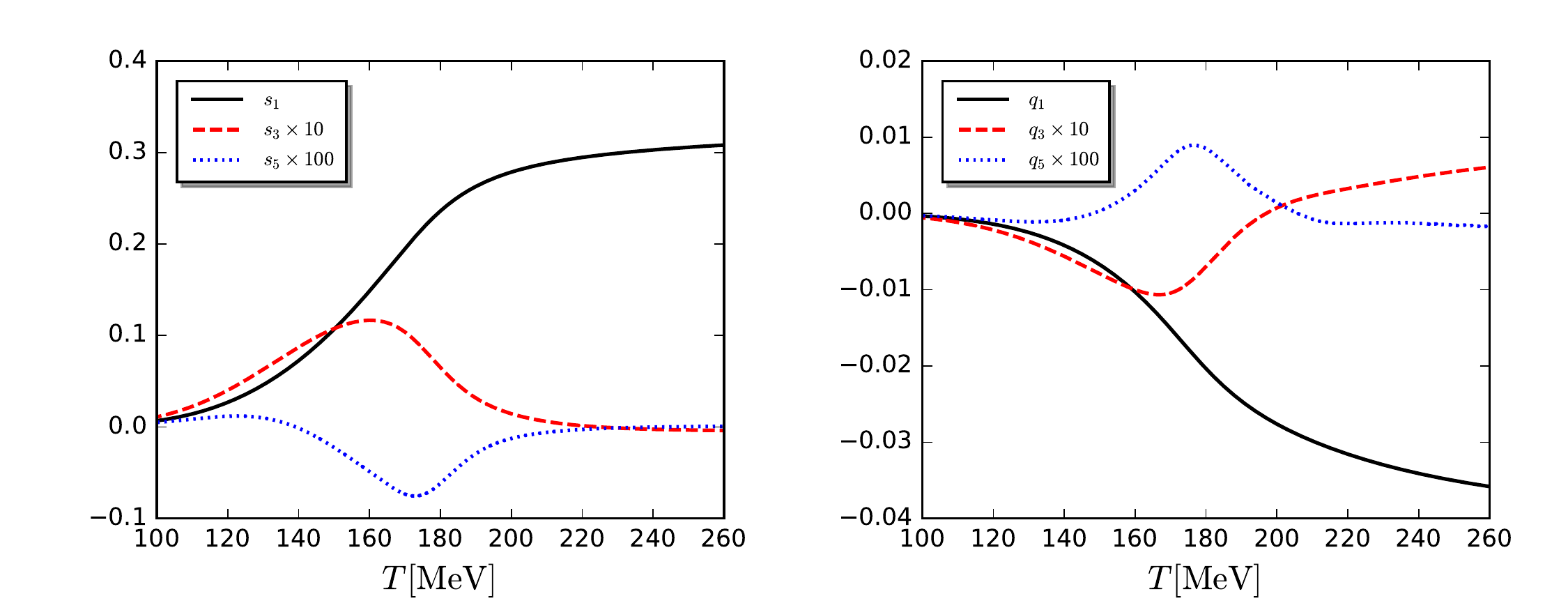}
\caption{Expansion coefficients of the strangeness chemical potential in \Eq{eq:muS0} (left panel) and electric charge chemical potential in \Eq{eq:muQ0}  (right panel), with $r=n_Q/n_B=0.4$ and $n_S=0$, as functions of the temperature.}\label{fig:snqn}
\end{figure*}
%

%
\begin{figure}[t]
\includegraphics[width=0.5\textwidth]{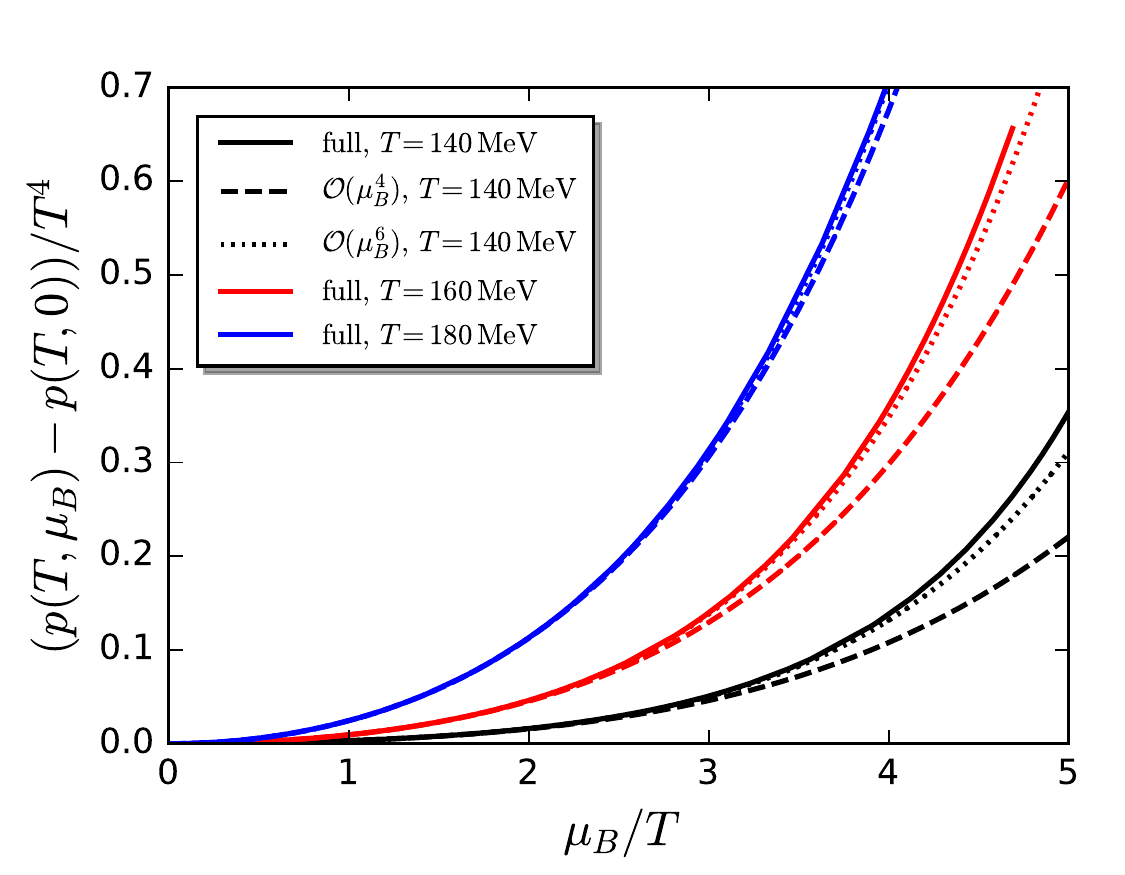}
\caption{Pressure subtracted by its value at $\mu_B=0$ as a function of the baryon chemical potential $\mu_B$ with $r=n_Q/n_B=0.4$ and $n_S=0$. The full results (solid lines) are compared with those from the Taylor expansion in \Eq{eq:pres2} up to the orders of $\mathcal{O}(\mu_B^4)$ (dashed lines) and $\mathcal{O}(\mu_B^6)$ (dotted lines). Different colors are used to denote different values of the temperature.}\label{fig:Pmu}
\end{figure}
%

The baryon number, electric charge, and the strangeness are conserved charges in heavy ion collisions, since they are not changed through the strong interactions of QCD. Thus, there are three relevant chemical potentials, i.e., $\mu_B$, $\mu_Q$ and $\mu_S$, which are related to the three-flavor quark chemical potentials through the relations as follow, $\mu_u=\mu_B/3 + 2\mu_Q/3$, $\mu_d=\mu_B/3-\mu_Q/3$, and $\mu_s=\mu_B/3 -\mu_Q/3 -\mu_S$. Here $\mu_u$, $\mu_d$, and $\mu_s$ are the chemical potentials for $u$, $d$, and $s$ quarks, respectively. When the chemical potentials are nonvanishing, the thermodynamical potential density $\Omega$, or the pressure $p$ in \Eq{eq:pressure} is a function of the temperature $T$ as well as the chemical potentials, to wit, $p(T, \mu_B,\mu_Q,\mu_S)$. Differentiating the pressure w.r.t the three different chemical potentials, one is led to the generalized susceptibilities, which read
\begin{align}
  \chi_{ijk}^{BQS}(T, \mu_B,\mu_Q,\mu_S)= \frac{\partial^{i+j+k} (p/T^4)}{\partial \hat \mu_B^i \partial \hat \mu_Q^j \partial \hat\mu_S^k}\,,\label{eq:chiBQS}
\end{align}
with $\hat \mu_X=\mu_X/T$ ($X=$$B$, $Q$, $S$). The generalized susceptibilities in \Eq{eq:chiBQS} are related to the cumulants of the conserved charge distributions, such as the diagonal ones:
\begin{align}
  \chi_1^X&=\frac{1}{VT^3}\big\langle N_X \big\rangle\,,\\[2ex]
  \chi_2^X&=\frac{1}{VT^3}\big\langle(\delta N_X)^2\big\rangle\,,\\[2ex]
  \chi_3^X&=\frac{1}{VT^3}\big\langle(\delta N_X)^3\big\rangle\,,\\[2ex]
  \chi_4^X&=\frac{1}{VT^3}\Big(\big\langle(\delta N_X)^4\big\rangle-3\big\langle(\delta N_X)^2\big\rangle^2\Big)\,,
\end{align}
with $\delta N_X=N_X-\langle N_X\rangle$, where the ensemble average is denoted by the symbol $\langle \cdots\rangle$. Here, $V$ is volume of the system, and one also often uses the number density $n_X=\langle N_X\rangle/V$. In the same way, one has similar relations for the off-diagonal cumulants of the conserved charge distributions, or the correlations among the conserved charges, e.g.,
\begin{align}
  \chi_{11}^{XY}&=\frac{1}{VT^3}\big\langle(\delta N_X) (\delta N_Y)\big\rangle\,,\\[2ex]
  \chi_{12}^{XY}&=\frac{1}{VT^3}\big\langle(\delta N_X) (\delta N_Y)^2\big\rangle\,,\\[2ex]
  \chi_{111}^{XYZ}&=\frac{1}{VT^3}\big\langle(\delta N_X) (\delta N_Y) (\delta N_Z)\big\rangle\,, \label{eq:chi111}
\end{align}
with $Y\,, Z \in \{B,\,Q,\,S\}$.

In the environment of heavy ion collision, the produced matter of quark-gluon plasma (QGP) is neutral in strangeness, which entails that the expected value of the density of strangeness is vanishing, i.e., $\langle n_S\rangle=0$. Furthermore, the ratio between the electric charge and the baryon number density $r=\langle n_Q\rangle/\langle n_B\rangle$ is a constant on average,  whose value is determined by the charge-mass ratio $Z/A$ of the colliding nuclei. In this work we adopt $r=0.4$ that is consistent with that of Au- or Pb- nucleus. Within these two constraints, one is led to the equations as follow
\begin{align}
  \chi_1^S(T, \mu_B,{\mu_Q}_0,{\mu_S}_0)&=0\,,\label{eq:chiS0}\\[2ex]
  \frac{\chi_1^Q(T, \mu_B,{\mu_Q}_0,{\mu_S}_0)}{\chi_1^B(T, \mu_B,{\mu_Q}_0,{\mu_S}_0)}&=r\,,\label{eq:chiQchiSr}
\end{align}
where we have defined ${\mu_Q}_0$ and ${\mu_S}_0$, which depend on $T$ and $\mu_B$. Employing Eqs. (\ref{eq:chiS0}) and (\ref{eq:chiQchiSr}), one can obtain the differential equations for ${\mu_Q}_0(T,\mu_B)$, ${\mu_S}_0(T,\mu_B)$ w.r.t. $\mu_B$, which read
\begin{align}
  \frac{\partial {\mu_{S}}_0(T,\mu_B)}{\partial \mu_B}&=-\frac{\chi^{BS}_{11}}{\chi^S_2}-\frac{\chi^{QS}_{11}}{\chi^S_2}\frac{\partial {\mu_Q}_0}{\partial \mu_B}\,,\label{eq:dmuSdmuB}\\[2ex]
\frac{\partial {\mu_Q}_0(T,\mu_B)}{\partial \mu_B}&=\frac{\chi^{BS}_{11}(\chi^{SQ}_{11}-r \chi^{BS}_{11})-\chi^S_2(\chi^{BQ}_{11}-r \chi^B_2)}
{\chi^S_2(\chi^Q_2-r \chi^{BQ}_{11})-\chi^{SQ}_{11}(\chi^{SQ}_{11}-r \chi^{BS}_{11})}\,.\label{eq:dmuQdmuB}
\end{align}
Equations (\ref{eq:dmuSdmuB}) and (\ref{eq:dmuQdmuB}) were first derived in \cite{Fu:2018swz}. These two differential equations, with together ${\mu_{S}}_0(T,0)={\mu_Q}_0(T,0)=0$, can be employed to obtain the values of ${\mu_{S}}_0$ and ${\mu_Q}_0$ at any finite $\mu_B$ for a fixed value of $T$. 

In \Fig{fig:muSQ0} we show the dependence of ${\mu_{S}}_0$ and ${\mu_Q}_0$ on the baryon chemical potential under the constraints $r=n_Q/n_B=0.4$ and $n_S=0$ at several values of temperature. One can see that the magnitude of ${\mu_Q}_0$ is quite smaller than that of ${\mu_{S}}_0$, which is reasonable, since ${\mu_Q}_0$ would be vanishing if the value of the ratio of electric charge to baryon number density is chosen to be $r=0.5$, due to the isospin symmetry between the $u$ and $d$ quarks. And in the environment of heavy ion collisions, $r=0.4$ is not far from the symmetric case. Concentrating on the left panel of \Fig{fig:muSQ0}, one finds that, when the temperature is high, ${\mu_{S}}_0$ approaches to the free quark gas limit ${\mu_{S}}_0=\mu_B/3$ as denoted by the black dashed line. However, this is not the case when the temperature is low and the system is near or below the chiral phase transition, where the deviation is remarkable. In summary, based on the results shown in \Fig{fig:muSQ0}, the commonly adopted approximation for the strangeness neutral matter produced in heavy ion collisions with $\mu_Q=0$ and $\mu_s=\mu_B/3-\mu_S=0$, is applicable in the chiral symmetric phase, but its validity should be treated carefully with the decrease of the temperature.

Note that the calculated ${\mu_{S}}_0$ and ${\mu_Q}_0$ in \Fig{fig:muSQ0} from \Eq{eq:dmuSdmuB} and \Eq{eq:dmuQdmuB} are exact and, in principle, can be extended to the regime of any large baryon chemical potential, which is in contradistinction to lattice QCD simulations, where the calculations are usually constrained to the region of $\mu_B/T\le 2\sim3$, due to the sign problem. In order to make better comparison between our results and those from lattice QCD, we adopt the approach of Taylor expansion in \cite{Bazavov:2017dus} to calculate the strangeness and electric charge chemical potentials with the constraints of the strangeness neutrality and the fixed ratio between the electric charge and the baryon number density. We begin with the pressure expanded around the vanishing chemical potentials, which reads
\begin{align}
  \frac{p(T, \mu_B,\mu_Q,\mu_S)}{T^4}&=\sum_{i,j,k=0}^\infty \frac{\chi_{ijk}^{BQS}(T, 0)}{ i!j!k!}\hat \mu_B^i \hat \mu_Q^j \hat\mu_S^k,\label{eq:pres}
\end{align}
where $\chi_{ijk}^{BQS}(T, 0)$ denotes the generalized susceptibility in \Eq{eq:chiBQS} with $\mu_B=\mu_Q=\mu_S=0$. In the same way, ${\mu_{S}}_0$ and ${\mu_Q}_0$ in Eqs. (\ref{eq:dmuSdmuB}) and (\ref{eq:dmuQdmuB}) can be expanded in a series of the baryon chemical potential as well, to wit,
\begin{align}
   {\hat \mu}_{S0}(T,\mu_B)&=s_1(T)\hat \mu_B+s_3(T){\hat \mu_B}^3+s_5(T){\hat \mu_B}^5+...,\label{eq:muS0}\\[2ex]
   {\hat \mu}_{Q0}(T,\mu_B)&=q_1(T)\hat \mu_B+q_3(T){\hat \mu_B}^3+q_5(T){\hat \mu_B}^5+...,\label{eq:muQ0}
\end{align}
where the coefficients $s_n$'s and $q_n$'s can be determined order by order, see \cite{Bazavov:2017dus} for more details. Inserting Eqs. (\ref{eq:muS0}) and (\ref{eq:muQ0}) into \Eq{eq:pres}, one is able to eliminate $\mu_S$ and $\mu_Q$ on the r.h.s., and thus the pressure is only expanded in terms of the baryon chemical potential, which reads
\begin{align}
  &\frac{p(T, \mu_B)}{T^4}-\frac{p(T, 0)}{T^4}\nonumber\\[2ex]
=&p_{2}(T){\hat \mu_B}^2+p_{4}(T){\hat \mu_B}^4+p_{6}(T){\hat \mu_B}^6+...\label{eq:pres2}
\end{align}

We show the leading order expansion coefficients $s_1$ and $q_1$ in Eqs. (\ref{eq:muS0}) and (\ref{eq:muQ0}) in Figure \ref{fig:S1Q1}, and the ratio of the next-leading to leading order coefficients, i.e., $s_3/s_1$ and $q_3/q_1$ in Figure \ref{fig:S31Q31}, which are also compared with the continuum extrapolated lattice results from the HotQCD Collaboration \cite{Bazavov:2017dus}. As we have discussed in our former work \cite{Wen:2018nkn}, the pseudo-critical temperatures at vanishing baryon chemical potential obtained from lattice simulations and effective models are usually not identical, due to some reasons, e.g., the different absolute scale. Accordingly, it is more appropriate to rescale the temperature for the $x$-axis in \Fig{fig:S1Q1} and \Fig{fig:S31Q31} by their respective $T_c$ at $\mu_B=0$. Note that the pseudo-critical temperature of the QCD chiral crossover at zero values of chemical potentials is $T_c=156$ MeV for lattice calculations by HotQCD Collaboration \cite{Bazavov:2018mes}. In our calculations we find $T_c^{\chi}=194$ MeV for the chiral crossover, and $T_c^{d}=177$ MeV for the deconfinement phase transition, see \cite{Wen:2018nkn} for more details and relevant discussions about these two pseudo-critical  temperatures. So, it is reasonable to adopt a value of $T_c$ in-between $T_c^{d}$ and $T_c^{\chi}$ for the fRG calculation. In \Fig{fig:S1Q1} and \Fig{fig:S31Q31} we employ $T_c=183$ MeV for  the fRG, which is motivated by the fact that, with this choice, the lattice and fRG show a consistent result for $s_1$ when the temperature is around and below $T_c$, as shown in the left plot of \Fig{fig:S1Q1}. It is, however, obvious that one can not match the calculated $s_1$ from the two different approaches in the whole temperature range shown in the plot, just by the tuning of $T_c$, and we observe that the fRG undershoots $s_1$ a bit as $T\gtrsim T_c$, in comparison to the lattice result. In the same way, from Figures \ref{fig:S1Q1} and \ref{fig:S31Q31} it is not difficult  to conclude that the fRG results are in qualitative, or even quantitative for some values of $T$, agreement with those of the lattice, e.g. both the two approaches find that the ratio $q_3/q_1$ crosses the zero line and changes sign at a value of $T$ larger than $T_c$, and $s_3/s_1$ is vanishing in the high temperature regime. Note however that the discrepancies between the results obtained from the fRG and the lattice are still of significance, for instance besides the leading order $s_1$ at high temperature, $|q_1|$ from fRG is also smaller than that from the lattice in the whole temperature region, as shown in the right panel of \Fig{fig:S1Q1}. Furthermore, one also finds that the next-leading to leading order ratios $s_3/s_1$ and $q_3/q_1$ obtained from the fRG are a bit smaller than those of the lattice in some regime of temperature.

As we have discussed above, due to the sign problem at finite density, the Taylor expansion is usually employed in lattice simulations, see e.g., Eqs. (\ref{eq:muS0}) (\ref{eq:muQ0}) (\ref{eq:pres2}). It is, therefore, of significant importance to investigate the convergency of the Taylor expansion in powers of $\mu_B/T$. In this work, we would like to study this convergency through the comparison between the Taylor expansion and full results. In \Fig{fig:muSQcompa} the full results of ${\mu_{S}}_0(T,\mu_B)$ and ${\mu_Q}_0(T,\mu_B)$ obtained from Eqs. (\ref{eq:dmuSdmuB}) and (\ref{eq:dmuQdmuB}) are compared with those from the Taylor expansions in Eqs. (\ref{eq:muS0}) and (\ref{eq:muQ0}), both of which are calculated within the fRG approach. Two results for the Taylor expansion are presented, which corresponds to the order of expansion up to $\mathcal{O}(\mu_B^3)$ and $\mathcal{O}(\mu_B^5)$, and are denoted by the dashed and dotted lines, respectively in \Fig{fig:muSQcompa}. Three different values of the temperature are adopted, and the relevant results are plotted in different colors. We find that the convergency of the Taylor expansion is observed for both the strangeness chemical potential and the electric charge chemical potential with $\mu_B/T\lesssim 2$; moreover, comparing these two chemical potentials, one can see the convergency property of ${\mu_S}_0$ is better than that of ${\mu_Q}_0$, and the Taylor expansion result of ${\mu_S}_0$ of order $\mathcal{O}(\mu_B^5)$ is still comparable to the full result when the $\mu_B$ is increased up to $\sim 3T$. Furthermore, in \Fig{fig:muSQcompa} one observes an alternating divergence with the increase of the order for $T=160$ and 180 MeV, while it is not for $T=140$ MeV. In order to explore the underlying mechanism, we show the lowest three coefficients of the Taylor expansion in Eqs. (\ref{eq:muS0}) and (\ref{eq:muQ0}) in \Fig{fig:snqn}, where they are rescaled appropriately for the convenience of presentation in one plot. One sees that when the temperature is low, $s_3$ and $s_5$ ($q_3$ and $q_5$) have the same signs, while with the increase of the temperature, they develops opposite signs. 

In \Fig{fig:Pmu} we show the dependence of the pressure on the baryon chemical potential within the constraints of the strangeness neutrality and the fixed ratio of the electric charge to baryon number density. We also compare the full results with those from the Taylor expansion. One observes that the convergency property for the pressure is better than that for the strangeness chemical potential and the electric charge chemical potential as shown in \Fig{fig:muSQcompa}, and the computation of the Taylor expansion up to order of $\mathcal{O}(\mu_B^6)$ is in good agreement with the full one, when the baryon chemical potential is increased to about four times the value of the temperature. In conclusion, the Taylor expansion of ${\mu_{S}}_0$ and ${\mu_Q}_0$ around $\mu_B=0$ up to the fifth order is found to be convergent with $\mu_B/T\lesssim 2\sim3$, and that of the pressure up to the sixth order is convergent with $\mu_B/T\lesssim 4$.

\section{Correlations of conserved charges}
\label{sec:correlation}

\begin{table}[t]
  \begin{center}
    \begin{tabular}{|c ||c|c|c|c|c|c|c|c|}
    \hline & & & & & & & & \\[-2ex]
    \hline & & & & & & & & \\[-1ex]
    $\sqrt{s_{NN}}$ [GeV] & 200 & 62.4 & 39 & 27 & 19.6 & 14.5 & 11.5 & 7.7\\[1ex]
    \hline & & & & & & & & \\[-2ex]
    ${\mu_B}_{_{CF}}^{^{LEFT}}$ [MeV] & 26 & 81 & 125 & 175 & 231 & 296 & 356 & 477\\[1ex]
    \hline & & & & & & & & \\[-2ex]
    $T_{_{CF}}^{^{LEFT}}$ [MeV] &186 & 186 & 185 & 184 & 183 & 179 & 175 & 162\\[1ex]
    \hline & & & & & & & & \\[-2ex]
    ${T_{c}^{\chi}}^{^{LEFT}}$ [MeV] & 193 & 193 & 192 & 191 & 190 & 187 & 184 & 175\\[1ex]
    \hline
    \end{tabular}
    \caption{Chemical freeze-out baryon chemical potential ${\mu_B}_{_{CF}}^{^{LEFT}}$ and temperature $T_{_{CF}}^{^{LEFT}}$ in the low energy effective theory for eight values of the collision energy $\sqrt{s_{NN}}$. In the last row ${T_{c}^{\chi}}^{^{LEFT}}$ is the pseudo-critical temperature of the chiral crossover at a fixed value of ${\mu_B}_{_{CF}}^{^{LEFT}}$, which is determined by the peak position of $|\partial \rho_1/\partial T|$ with $\rho_1$ given in \Eq{eq:rho1}.}
    \label{tab:freeze-out-para}  
  \end{center}\vspace{-1cm}
\end{table}

%
\begin{figure*}[t]
\includegraphics[width=1.\textwidth]{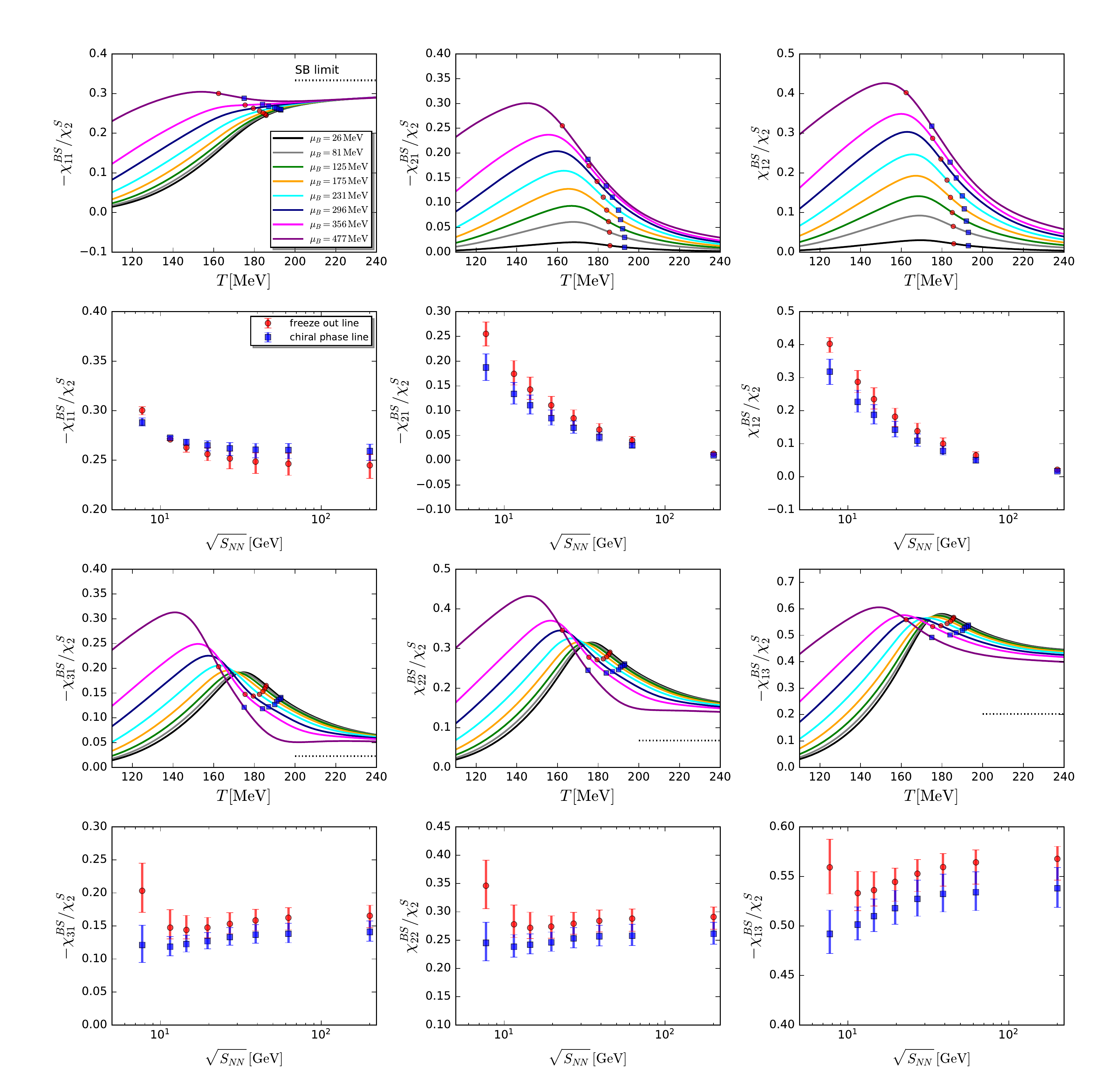}
\caption{Correlations between the baryon number and the strangeness of orders from the second to the fourth, as functions of the temperature (subplots in the first and third rows) at values of the chemical freeze-out baryon chemical potential ${\mu_B}_{_{CF}}^{^{LEFT}}$ in \tab{tab:freeze-out-para}, where the two constraints, i.e., the strangeness neutrality $n_S=0$ and a fixed value of the ratio between the electric charge and the baryon number density $r=n_Q/n_B=0.4$ are implemented. On the different lines related to the different chemical potentials or different values of the collision energy, we use the red circles to denote the freeze-out temperature $T_{_{CF}}^{^{LEFT}}$ and the blue squares to the chiral pseudo-critical temperature ${T_{c}^{\chi}}^{^{LEFT}}$ in \tab{tab:freeze-out-para}. The two sets of points are also plotted as functions of the collision energy in an associating subplot for every subplot above it. In the subplots of the collision energy, we use error bars to indicate the variation of correlations at the temperature $T_{_{CF}}^{^{LEFT}}\pm 5$ MeV and ${T_{c}^{\chi}}^{^{LEFT}}\pm 5$ MeV.}\label{fig:chiBS}
\end{figure*}
%

%
\begin{figure*}[t]
\includegraphics[width=1.\textwidth]{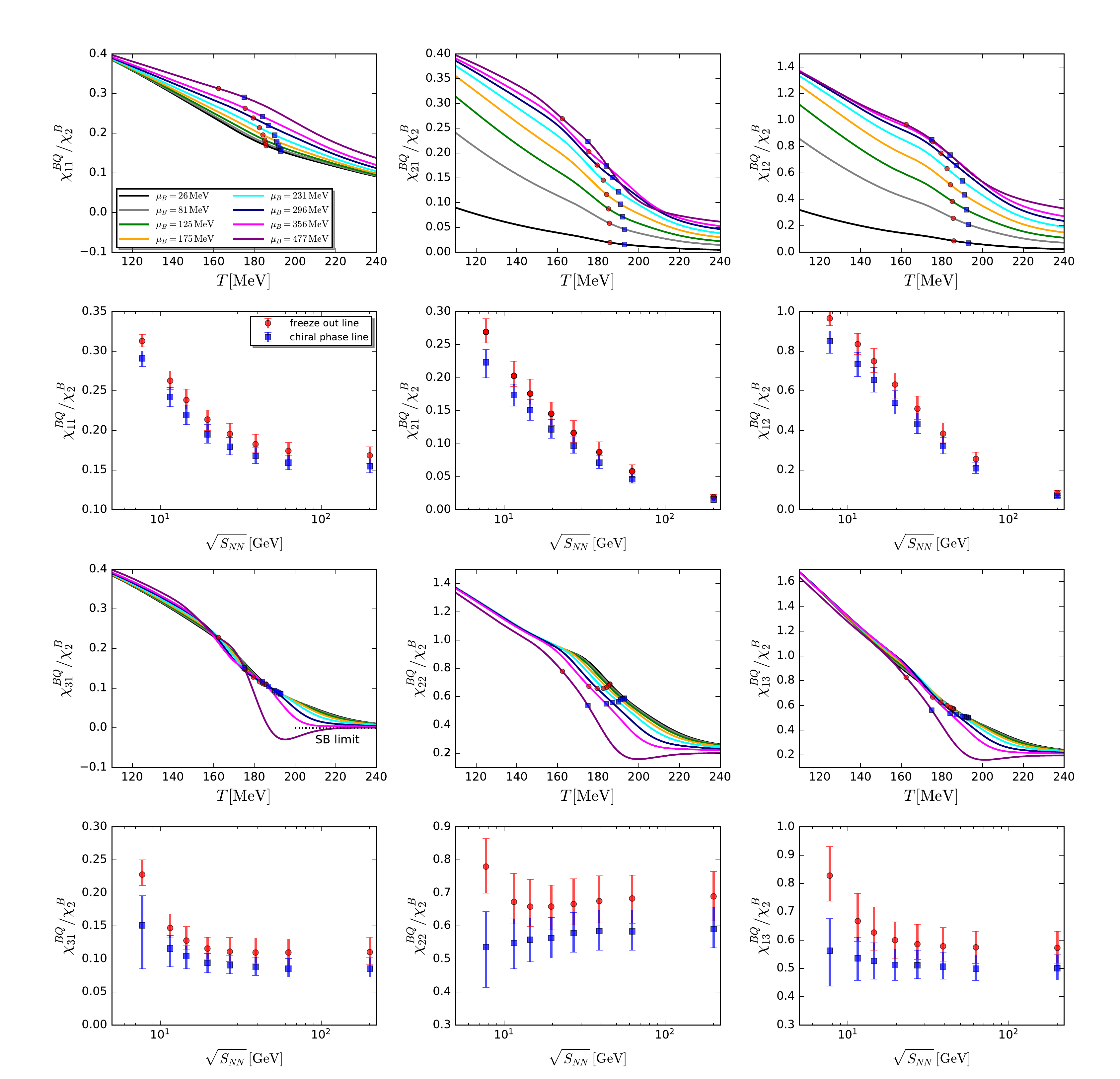}
\caption{Correlations between the baryon number and the electric charge of orders from the second to the fourth, as functions of the temperature (subplots in the first and third rows) at values of the chemical freeze-out baryon chemical potential ${\mu_B}_{_{CF}}^{^{LEFT}}$ in \tab{tab:freeze-out-para}, where the two constraints, i.e., the strangeness neutrality $n_S=0$ and a fixed value of the ratio between the electric charge and the baryon number density $r=n_Q/n_B=0.4$ are implemented. On the different lines related to the different chemical potentials or different values of the collision energy, we use the red circles to denote the freeze-out temperature $T_{_{CF}}^{^{LEFT}}$ and the blue squares to the chiral pseudo-critical temperature ${T_{c}^{\chi}}^{^{LEFT}}$ in \tab{tab:freeze-out-para}. The two sets of points are also plotted as functions of the collision energy in an associating subplot for every subplot above it. In the subplots of the collision energy, we use error bars to indicate the variation of correlations at the temperature $T_{_{CF}}^{^{LEFT}}\pm 5$ MeV and ${T_{c}^{\chi}}^{^{LEFT}}\pm 5$ MeV.}\label{fig:chiBQ}
\end{figure*}
%

%
\begin{figure*}[t]
\includegraphics[width=1.\textwidth]{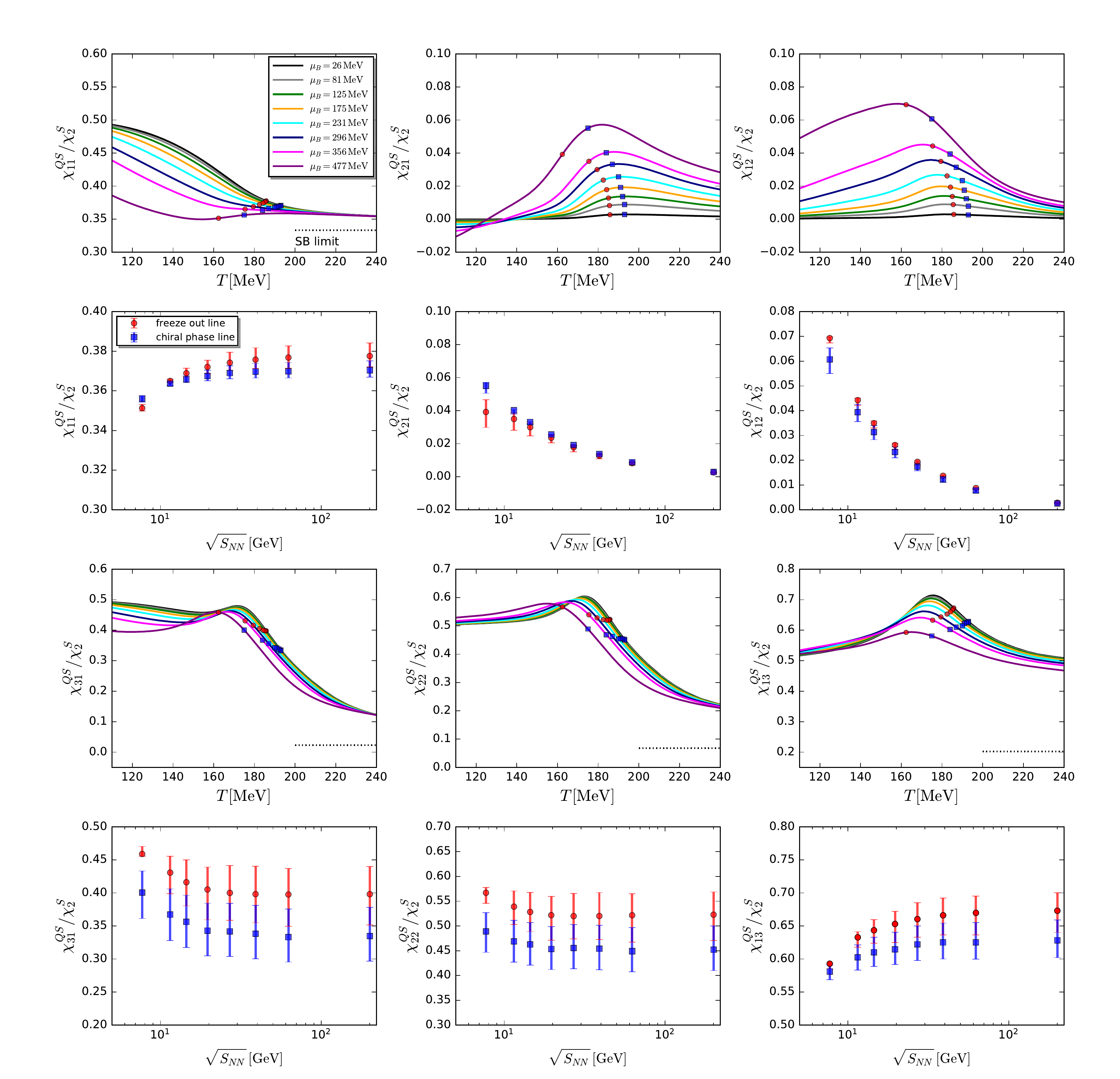}
\caption{Correlations between the electric charge and the strangeness of orders from the second to the fourth, as functions of the temperature (subplots in the first and third rows) at values of the chemical freeze-out baryon chemical potential ${\mu_B}_{_{CF}}^{^{LEFT}}$ in \tab{tab:freeze-out-para}, where the two constraints, i.e., the strangeness neutrality $n_S=0$ and a fixed value of the ratio between the electric charge and the baryon number density $r=n_Q/n_B=0.4$ are implemented. On the different lines related to the different chemical potentials or different values of the collision energy, we use the red circles to denote the freeze-out temperature $T_{_{CF}}^{^{LEFT}}$ and the blue squares to the chiral pseudo-critical temperature ${T_{c}^{\chi}}^{^{LEFT}}$ in \tab{tab:freeze-out-para}. The two sets of points are also plotted as functions of the collision energy in an associating subplot for every subplot above it. In the subplots of the collision energy, we use error bars to indicate the variation of correlations at the temperature $T_{_{CF}}^{^{LEFT}}\pm 5$ MeV and ${T_{c}^{\chi}}^{^{LEFT}}\pm 5$ MeV.}\label{fig:chiQS}
\end{figure*}
%

In this section we would like to investigate the correlations of conserved charges at finite temperature and density, in the 2+1 flavor low energy effective theory within the fRG approach. Two constraints, i.e., the strangeness neutrality $n_S=0$ and a fixed value of the electric charge to baryon number density $r=n_Q/n_B=0.4$ are implemented. We focus on the correlations between any two different conserved charges, i.e., those of form $\chi_{n_{_{X}}n_{_{Y}}}^{XY}$ with $X\,, Y \in \{B,\,Q,\,S\}$. All the relevant correlations of orders from the quadratic to quartic, viz., $n_{_{X}}+n_{_{Y}}=$2, 3, and 4, have been calculated. The motivations of this extensive study are given as follows. On the one hand, from the theoretical side,  it is valuable to study the correlations, in particular higher-order ones and in the circumstance of heavy-ion collisions with strangeness neutrality and a fixed value of the electric charge to baryon number density, in the fRG approach. And the relevant calculated results are easily compared with other theoretical calculation, e.g., the lattice results \cite{Bellwied:2019pxh}. On the other hand, from the experimental side, through an extensive study we would like to provide some useful information for experimentalists who are searching for the CEP. The dependence of various correlations on the temperature, baryon chemical potential, and the collision energy have been investigated, and several correlations which manifest the most non-monotonic behavior have been picked, which would be potentially useful in future experiments.

In fact, the second-order correlations between the net proton, net charge, and the net kaon have been measured by STAR Collaboration \cite{Adam:2019xmk}. Although these observables are related to $\chi^{BS}_{11}$, $\chi^{BQ}_{11}$ and $\chi^{QS}_{11}$ investigated in this work, it is still challenging to compare the experimental measurements with the theoretical calculations directly. It is of particular interest to study the relations among them, see, e.g., \cite{Bluhm:2020mpc} for related discussions.

Though till now only the quadratic correlations have been measured in the experiments, higher-order correlations as same as the higher-order fluctuations, see e.g.\cite{Adamczyk:2013dal,Adamczyk:2014fia,Luo:2015ewa}, carry more sensitive information on the critical dynamics of the chiral symmetry, for instance the non-monotonic behavior observed for the kurtosis of the net proton distribution as a function of the collision energy, for more details, see e.g. \cite{Luo:2017faz} and references therein. Therefore, it is more interesting to study the higher-order correlations in heavy ion collisions from the theoretical side, especially their dependence on the collision energy and potentially possible non-monotonic behaviors, which might provide some useful insights for the experimental measurements of the higher-order correlations in the future. Nonetheless, it is cautioned that our results should be taken with a grain of salt, since many effects, such as the non-equilibrium evolution of the system \cite{Rajagopal:2019xwg}, centrality and rapidity dependence, volume fluctuations, resonance decays \cite{Li:2017via}, etc. have not been taken into account in our studies, and they may make remarkable influences on a direct confrontation of the theoretical prediction with experimental measurements, see e.g., \cite{Bluhm:2020mpc} for a relevant review.

In order to investigate the dependence of conserved charge correlations on difference values of the collision energy, 
we employ the chemical freeze-out (CF) temperature $T_{_{CF}}$ and the baryon chemical potential ${\mu_B}_{_{CF}}$ in \cite{Andronic:2017pug}, which are obtained from the statistical hadronization approach. The parametrization for the dependence of $T_{_{CF}}$ and ${\mu_B}_{_{CF}}$ on the collision energy $\sqrt{s_{NN}}$ reads
\begin{align}
   T_{_{CF}}&=\frac{T_{_{CF}}^{lim}}{1+\exp\Big(2.60-\ln\big(\sqrt{s_{NN}}\big)/0.45\Big)},\label{}
\end{align}
and
\begin{align}
   {\mu_B}_{_{CF}}&=\frac{a}{1+0.288\sqrt{s_{NN}}},\label{}
\end{align}
with $T_{_{CF}}^{lim}=158.4$ MeV, $a=1307.5$ MeV, see e.g.,\cite{Karsch:2010ck} for more details about the freeze-out conditions in heavy ion collisions. Moreover, as we have discussed above, the scale in the low energy effective theory is a bit different from that in QCD, and this mismatch is taken into account for the freeze-out parameters through the rescale as follows
\begin{align}
   T_{_{CF}}^{^{LEFT}}&=\beta\,T_{_{CF}},\quad \mathrm{and} \quad {\mu_B}_{_{CF}}^{^{LEFT}}=\beta\,{\mu_B}_{_{CF}}\,,\label{}
\end{align}
where the quantities with the superscript $LEFT$ denote those in the low energy effective theory, and $\beta$ is a rescale coefficient. We employ the pseudo-critical temperature from the lattice simulation ${T_c}^{^{LAT}}=156$ MeV at vanishing chemical potential, and ${T_c}^{^{LEFT}}=183$ MeV in our calculation in the low energy effective theory, as discussed in \sec{sec:Thermo}, to determine the rescale coefficient $\beta$, which yields
\begin{align}
   \beta&=\frac{{T_c}^{^{LEFT}}}{{T_c}^{^{LAT}}}\simeq 1.173\,.\label{eq:beta}
\end{align}
Within this setup, values of the chemical freeze-out parameters in the low energy effective theory ${\mu_B}_{_{CF}}^{^{LEFT}}$ and $T_{_{CF}}^{^{LEFT}}$ are collected in \tab{tab:freeze-out-para}, corresponding to the eight values of the collision energy. Furthermore, we have also investigated the dependence of calculated results on the approach for the determination of the rescale coefficient $\beta$. We employ ${T_c}^{^{LEFT}}={T_{c}^{\chi}}^{^{LEFT}}=194$ MeV and ${T_c}^{^{LEFT}}=T_c^d=177$ MeV in lieu of ${T_c}^{^{LEFT}}=183$ in \Eq{eq:beta}, which corresponds to $\beta=({T_c}^{^{LEFT}}=194)/({T_c}^{^{LAT}}=156)\simeq 1.244$ and $\beta=({T_c}^{^{LEFT}}=177)/({T_c}^{^{LAT}}=156)\simeq 1.135$, respectively, and repeat all the calculations in Figs. \ref{fig:chiBS}, \ref{fig:chiBQ}, \ref{fig:chiQS}. We find that various correlations as functions of the collision energy are not changed qualitatively. In order to investigate quantitative errors resulting from the freeze-out temperature further, we also consider the case that the freeze-out temperature coincides with the chiral pseudo-critical temperature ${T_{c}^{\chi}}^{^{LEFT}}$ for every value of ${\mu_B}_{_{CF}}^{^{LEFT}}$, or the collision energy $\sqrt{s_{NN}}$, and the relevant values of ${T_{c}^{\chi}}^{^{LEFT}}$ are listed in the last row of in \tab{tab:freeze-out-para}.

Numerical results of the correlations $\chi^{BS}_{n_{_{B}}n_{_{S}}}$, $\chi^{BQ}_{n_{_{B}}n_{_{Q}}}$ and $\chi^{QS}_{n_{_{Q}}n_{_{S}}}$ are shown in figures \ref{fig:chiBS}, \ref{fig:chiBQ}, and \ref{fig:chiQS}, respectively. They are normalized by either the quadratic fluctuation of the strangeness $\chi^{S}_{2}$ or that of the baryon number $\chi^{B}_{2}$. In every figure, six different correlations of the second order $\chi^{XY}_{11}$, the third order $\chi^{XY}_{21}$, $\chi^{XY}_{12}$, and the fourth order $\chi^{XY}_{31}$, $\chi^{XY}_{22}$, $\chi^{XY}_{13}$ are presented. They are depicted as functions of the temperature with different values of ${\mu_B}_{_{CF}}^{^{LEFT}}$ in \tab{tab:freeze-out-para}, in the subplots of the first and third rows. We also calculate the Stefan-Boltzmann (SB) limit values of the three-flavor massless free quark gas for the various correlations analytically. In order to simplify the calculations of the SB limits, one has assumed that the electric charge chemical potential is vanishing, i.e., ${\mu_Q}_0=0$, which is a reasonable approximation given its small value as shown in the right panel of \Fig{fig:muSQ0}. Then we have ${\mu_{S}}_0=\mu_B/3$, demanded by the strangeness neutrality. We plot the SB values for some selective correlations in Figs. \ref{fig:chiBS}, \ref{fig:chiBQ}, \ref{fig:chiQS} with the black dotted lines, for which the SB values are constant and do not depend on the baryon chemical potential. Notice that for $\chi^{QS}_{31}/\chi^{S}_{2}$, $\chi^{QS}_{22}/\chi^{S}_{2}$ and $\chi^{QS}_{13}/\chi^{S}_{2}$ in \Fig{fig:chiQS}, one observes that the deviation of the calculated values from their SB values in the high temperature regime grows with the increasing order $n_{_{S}}$ from 1 to 3. This is due to the fact that quantum fluctuations of open strange mesons, such as kaons, still play a significant role in this region in the low energy effective theory, see relevant discussions in \cite{Wen:2018nkn} for more details, which is an indication that the low energy effective theory should be improved toward the description of QCD, see e.g. \cite{Fu:2019hdw}, where the mesonic degrees of freedom are quickly decoupled, once the temperature is above the pseudo-critical one.

Employing the chemical freeze-out temperature $T_{_{CF}}^{^{LEFT}}$ and the chiral pseudo-critical temperature ${T_{c}^{\chi}}^{^{LEFT}}$ for the eight values of the collision energy in \tab{tab:freeze-out-para}, we plot all the correlations as functions of the collision energy in the second and fourth rows of Figures \ref{fig:chiBS}, \ref{fig:chiBQ}, \ref{fig:chiQS}, each below their respective subplots showing the $T$-dependence. The correlations determined from $T_{_{CF}}^{^{LEFT}}$ and ${T_{c}^{\chi}}^{^{LEFT}}$ are denoted with the red circles and blue squares, legended with the freeze-out line and chiral phase line, respectively. And their positions are also shown in the $T$-subplots. Furthermore, in order to estimate the errors arising from the determinations of $T_{_{CF}}^{^{LEFT}}$ and ${T_{c}^{\chi}}^{^{LEFT}}$, we employ the error bars to represent the values of correlations at the temperature $T_{_{CF}}^{^{LEFT}}\pm 5$ MeV and ${T_{c}^{\chi}}^{^{LEFT}}\pm 5$ MeV.

From Figs. \ref{fig:chiBS}, \ref{fig:chiBQ}, \ref{fig:chiQS}, one can see that higher-order correlations are more interesting than the second-order ones, since the singular part, that contributes to the various correlations and is related to the critical dynamics, holds increasing importance with the increase of the order of correlations. This is confirmed by the dependence of all the correlations on the temperature with a fixed value of the baryon chemical potential, where non-monotonic behaviors become more and more obvious with the increasing order. Of all the correlations presented in Figs. \ref{fig:chiBS}, \ref{fig:chiBQ}, \ref{fig:chiQS}, one observes that two of the fourth-order correlations between the baryon number and the strangeness, $-\chi^{BS}_{31}/\chi^{S}_{2}$ and $\chi^{BS}_{22}/\chi^{S}_{2}$ manifest themselves as the most non-monotonic correlations, as shown pronouncedly in the third row of Fig. \ref{fig:chiBS}, where they are plotted as functions of temperature. Moreover, it is found that the dependence of these two fourth-order correlations on the collision energy is non-monotonic as well, both for the freeze-out line and the chiral phase line. One observes that these two correlations decrease a bit when the collision energy decreases from the 200 GeV to 14.5 GeV, and then goes up after the collision energy is reduced further. In short summary, after an extensive study of all the correlations of conserved charges up to the fourth order, we find two of them, i.e., those between the baryon number and the strangeness, $-\chi^{BS}_{31}/\chi^{S}_{2}$ and $\chi^{BS}_{22}/\chi^{S}_{2}$, have the most significant non-monotonic behaviors, which would be potentially useful in the BES program at RHIC to search for the CEP.


\section{Summary and discussions}
\label{sec:sum}

In this work, we have studied the correlations of conserved charges, i.e., the baryon number, electric charge and the strangeness, up to the fourth order at nonzero temperature and chemical potentials. The computations are performed in a 2+1 flavor low energy effective theory within the fRG approach, following the setup presented in our former work \cite{Wen:2018nkn}. Employing a set of phenomenological chemical freeze-out parameters, we also study the possible evolution behavior of the correlations with the collision energy in the beam energy scan experiments at RHIC.

Computations in this work are performed for strangeness neutral systems with a fixed ratio of the electric charge to baryon number density, which mimic the environment in the heavy ion collisions. Due to these two constraints, the equilibrium strangeness and electric charge chemical potentials, denoted in this work by ${\mu_{S}}_0$ and ${\mu_Q}_0$, respectively, develop dependence on the baryon chemical potential $\mu_B$, which have been calculated fully in our approach. Furthermore, we also expand ${\mu_{S}}_0$ and ${\mu_Q}_0$ in powers of $\mu_B/T$, and the leading and next-leading order coefficients are compared with the lattice results. We find that, although there is a bit of quantitative deviation, the fRG results are in qualitative agreement with those of the lattice QCD.  The Taylor expansion of ${\mu_{S}}_0$ and ${\mu_Q}_0$ around $\mu_B=0$ up to the fifth order is found to be convergent with $\mu_B/T\lesssim 2\sim3$, and the convergency property for the pressure is relatively better with $\mu_B/T\lesssim 4$.

The dependence of various correlations on the temperature and baryon chemical potential have been investigated, and we find higher-order correlations, in comparison to the quadratic ones, are more sensitive to the critical dynamics. Of all the correlations of conserved charges up to the fourth order, we find two of them, the fourth-order correlations between the baryon number and strangeness normalized by the variance of strangeness, $-\chi^{BS}_{31}/\chi^{S}_{2}$ and $\chi^{BS}_{22}/\chi^{S}_{2}$, manifest themselves as the most non-monotonic correlations, and they also have a non-monotonic dependence on the collision energy. But we should emphasized that this result should be taken with a grain of salt. On the one hand, as we have discussed above, many effects, especially the non-equilibrium evolution of the system have not yet been taken into account in our calculations, and they may play a significant role in the comparison between the theoretical prediction and the experimental measurements; on the other hand, in the experimental measurements of correlations, net-proton and net-kaon multiplicity distributions are used as proxies for those of net baryon and net strangeness, respectively. However, the relations among them certainly need to be elucidated more clearly, which are delayed to future work.


\begin{acknowledgments}

We thank the members of the fQCD collaboration \cite{fQCD} for work on related projects. The work was supported by the National Natural Science Foundation of China under Contracts Nos. 11775041.

\end{acknowledgments}


\bibliography{ref-lib.bib}

\end{document}